\documentclass[12pt]{article}
 \usepackage{graphicx}

\makeatletter

\def\paragraph{\@startsection{paragraph}{4}{\z@}{+2.00ex plus
 +1ex minus +.2ex}{1.5ex plus .2ex}{\it\normalsize}}

\def\section{\@startsection {section}{1}{\z@}{+3.0ex plus +1ex minus
  +.2ex}{2.3ex plus .2ex}{\normalsize\bf\boldmath}}
\def\subsection{\@startsection{subsection}{2}{\z@}{+2.5ex plus +1ex
minus +.2ex}{1.5ex plus .2ex}{\normalsize\bf\boldmath}}
\def\subsubsection{\@startsection{subsubsection}{3}{\z@}{+3.25ex plus
 +1ex minus +.2ex}{1.5ex plus .2ex}{\normalsize\bf\boldmath}}

\expandafter\ifx\csname mathrm\endcsname\relax\def\mathrm#1{{\rm #1}}\fi

\newcounter{saveeqn}

\@addtoreset{equation}{section}

\def\PLB{{ Phys. Lett.}  B }

\def\amu{a_\mu}

\def\MZ{M_Z}

\def\dalf{\Delta\alpha}

\def\dah{\Delta\alpha^{(5)}_{\rm had}}

\def\dahz{\Delta\alpha^{(5)}_{\rm had}(\MZ^2)}
\def\dahzE{\Delta\alpha^{(5)}_{\rm had}(-\MZ^2)}
\def\dah0{\Delta\alpha^{(5)}_{\rm had}(-s_0)}
\def\sf{spectral function}
\def\sfs{spectral functions}

\newcommand{\sigh}{$\sigma(\epm \ra hadrons)$ }

\newcommand{\gv}{\mbox{GeV}}
\newcommand{\mv}{\mbox{MeV}}

\newcommand{\MOM}{${\mathrm{MOM}}$ }

\newcommand{\MSb}{$\overline{\mathrm{MS}}$ }

\newcommand{\MSbm}{\overline{\mathrm{MS}} }
\newcommand{\al }{\alpha}
\newcommand{\epm}{e^+e^-}

\newcommand{\be}{\begin{equation}}
\newcommand{\ee}{\end{equation}}
\newcommand{\ba}{\begin{eqnarray}}
\newcommand{\ea}{\end{eqnarray}}
\newcommand{\bea}{\begin{eqnarray*}}
\newcommand{\eea}{\end{eqnarray*}}
\newcommand{\bet}{\begin{center} \begin{tabular}}
\newcommand{\ent}{\end{tabular} \end{center}}
\newcommand{\bb}{\begin{thebibliography}}
\newcommand{\eb}{\end{thebibliography}}

\newcommand{\ra}{\rightarrow}

\newcommand{\bit}{\begin{itemize}}
\newcommand{\eit}{\end{itemize}}

\newcommand{\noi}{\noindent}

\newcommand{\lapprox}{\raisebox{-.2ex}{$\stackrel{\textstyle<}
{\raisebox{-.6ex}[0ex][0ex]{$\sim$}}$}}

\newcommand{\crn}{\nonumber \\}
\newcommand{\nn}{\nonumber}
\newcommand{\ha}{\frac{1}{2}}
\newcommand{\dal}{\Delta \alpha}

\newcommand{\mz}{M^2_Z}

\newcommand{\sinf}{\sin^2 \Theta_f}
\newcommand{\cosf}{\cos^2 \Theta_f}
\newcommand{\sini}{\sin^2 \Theta_i}
\newcommand{\cosi}{\cos^2 \Theta_i}
\newcommand{\sinW}{\sin^2 \Theta_W}
\newcommand{\cosW}{\cos^2 \Theta_W}
\newcommand{\sing}{\sin^2 \Theta_g}

\newcommand{\dro}{\Delta \rho}
\newcommand{\Gmu}{G_{\mu}}
\newcommand{\bary}{\begin{array}}
\newcommand{\eary}{\end{array}}

\newcommand{\damt}{\frac{\delta m_t}{m_t}}

\newcommand{\gmunuc}{g_{\mu \nu}}
\newcommand{\ep}{\;\: .}

\renewcommand{\thefootnote}{\fnsymbol{footnote}}
\begin{document}
%=====================================================================
\vspace*{-1cm}
\begin{flushright}
{\normalsize \rm
DESY 03-106\\
SFB/CPP-03-17\\
hep-ph/0308117\\
August 11, 2003
\vspace*{0.5cm}}
\end{flushright}

\begin{center}

{\Large \bf Hadronic vacuum polarization effects in $\alpha_{\rm
em}(M_Z)$\footnote[1]{\noindent \it To appear in the Proceedings of the
Mini-Workshop {\sc ``ELECTROWEAK PRECISION DATA AND THE HIGGS MASS''}
DESY Zeuthen, February 28th to March 1st, 2003 }\footnote[2]{Work supported
in part by TMR, EC-Contract No.~HPRN-CT-2002-00311 (EURIDICE), and by
DFG under Contract SFB/TR 9-03.} }

\vspace*{1cm}

{\sc F. Jegerlehner}

\vspace*{.5cm}

{\normalsize \it
DESY \\
Platanenallee 6, D-15738 Zeuthen, Germany}

\par
\end{center}
\vskip 1cm
\begin{center}
\bf Abstract
\end{center}
{\it Recent evaluations of the hadronic vacuum polarization
contributions to the effective fine-structure constant $\alpha_{\rm
em}(M_Z)$ are summarized and commented. A new update based on
corrected CMD-2 data is presented. My new estimates are $\Delta \al
_{\rm had}^{(5)}(\mz) = 0.027773 \pm 0.000354$ ($\epm$--data based)
and $\Delta \al _{\rm had}^{(5)}(\mz) = 0.027664 \pm 0.000173$ (via
the Adler-function and extended use of pQCD). Prospects of further
possible progress is discussed.}
\par
\vskip 6mm

\renewcommand{\thefootnote}{\arabic{footnote}}
\setcounter{footnote}{0}

\section{Introduction}

Precision physics requires appropriate inclusion of higher order
effects and the knowledge of very precise input parameters of the
electroweak Standard Model SM. One of the basic input parameters is
the fine structure constant which depends logarithmically on the
energy scale. Vacuum polarization effects lead to a partial screening
of the charge in the low energy limit (Thomson limit) while at higher
energies the strength of the electromagnetic interaction grows.
Non-perturbative strong interaction effects (virtual hadron
fluctuations) lead to a partially non-perturbative relationship
between the very precisely known classical fine structure constant
$\alpha$ and its effective value $\alpha(\mu^2)$ at nonzero energy
scales $\mu$. Presently, the only save way to evaluate the
non-perturbative hadronic contributions is via a dispersion integral
over experimental $\epm \to $ hadrons data (see
e.g.~\cite{FJ86,EJ95}).  The required relationship derives from
analyticity (as a consequence of causality) and the optical theorem
(unitarity). A drawback is that the experimental errors allow us to
calculate the shift in the fine structure constant only at limited
accuracy. In fact, at present, the corresponding uncertainty imposes
one of the limiting factors in making precise SM predictions. While
$\delta \dal$ is dominating the uncertainty $\delta \sinf$ the
experimental error of the top mass measurement $\delta m_t$ is the
main uncertainty contributing to $\delta M_W$. Both measurements
$\sinf$ and $M_W$ yield constraints on the as yet unknown Higgs
mass $M_H$.

The photon propagator provides a simple way to derive the
concept of an effective charge. Including a factor $e^2$ and
considering the renormalized ( multiplied by the appropriate wave
function renormalization factor $Z$) full photon propagator we have
\be
e^2D_{\mu\nu}(q)=\frac{\gmunuc \: e^2\:Z}{q^2\:\left( 1+\Pi'_\gamma(q^2)\right)}+ {\rm \
gauge \ terms \ }
\ee
which in effect means that the charge has to be replaced by a {\em
running charge}
\be
e^2 \to e^2(q^2)=\frac{e^2Z}{1+\Pi'_\gamma(q^2)} \ep
\ee
The wave function renormalization factor $Z$ is fixed by the condition
that for $q^2 \to 0$ one obtains the classical charge (charge
renormalization in the Thomson limit). Thus the renormalized charge is
\be
e^2 \to e^2(q^2)=\frac{e^2}{1+(\Pi'_\gamma(q^2)-\Pi'_\gamma(0))}
\label{runninge}
\ee
where, in perturbation theory, the lowest order diagram which
contributes to $\Pi'_\gamma(q^2)$ is

\centerline{%
\includegraphics[scale=0.65]{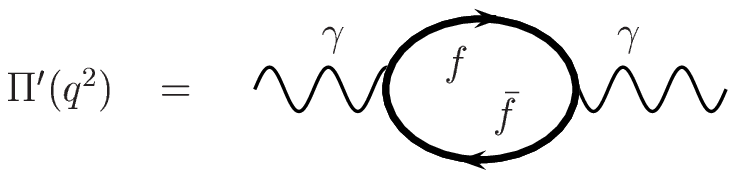}
}
\noi which describes virtual creation and reabsorption of
fermion pairs $${\gamma^*} \rightarrow {e^+e^-,\mu^+ \mu^-,
\tau^+ \tau^-, u\bar{u}, d
\bar{d},} \ {\cdots} {\rightarrow} {~~\gamma^*}$$
\noindent
in leading order.  In terms of the fine structure constant
$\alpha=e^2/4\pi$ Eq.~({\ref{runninge}) reads
\be
\alpha(q^2)=\frac{\alpha}{1-\Delta \alpha}\;\;\;;\;\;\;\Delta \alpha =
- {\rm Re}\:\left(\Pi'_\gamma(q^2)-\Pi'_\gamma(0)\right) \ep
\label{VPamp}
\ee

The shift $\Delta \alpha$ is large due to the large change in scale
going from zero momentum to the Z-mass scale $\mu=M_Z$ and due to the
many species of fermions contributing. Zero momentum more precisely
means the light fermion mass thresholds.

The various contributions to the shift in the fine structure constant
come from the leptons (lep = $e$, $\mu$ and $\tau$) the 5 light quarks
($u$, $b$, $s$, $c$, and $b$ and the corresponding hadrons = had) and
from the top quark:
\be
\Delta \alpha =\Delta \alpha_{\rm lep}+\Delta^{(5)} \alpha_{\rm had}
+\Delta \alpha_{\rm top}+\cdots
\ee
Also $W$--pairs contribute at $q^2 > 2 M_W^2$ (see~\cite{JF85,FJLCN}).
The leptonic contributions are calculable in perturbation theory where
at leading order the free lepton loops yield
\be \bary{l}
\dalf_{{\rm lep}}(s)= \cr \bary{lcl}
& = & \sum\limits_{\ell=e,\mu,\tau}
      \frac{\alpha}{3\pi}
      \left[ - \frac{8}{3} + \beta_\ell^2
             - \frac{1}{2}\beta_\ell(3 - \beta_\ell^2)
               \ln\left( \frac{1-\beta_\ell}{1+\beta_\ell}
                  \right)
      \right]   \cr
& = &
      \sum\limits_{\ell=e,\mu,\tau}
      \frac{\alpha}{3\pi}
      \left[ \ln\left( s/m_\ell^2
                \right)
           - \frac{5}{3}
           + O\left( m_\ell^2/s
              \right)
      \right]  {\rm \ for \ } |s|\gg m_\ell^2   \cr
& \simeq & 0.03142 {\rm \ for \ } s=M_Z^2 \eary \eary
\ee
where $\beta_\ell = \sqrt{1 - 4m_\ell^2/s}$. This leading contribution
is affected by small electromagnetic corrections only in the next to
leading order. The leptonic contribution is actually known to three
loops~\cite{KalSab55,Ste98} at which it takes the value ($M_Z \sim
91.19$ GeV)\footnote{For $m_t \sim 174.3$ GeV we have $\Delta \alpha_{\rm
top}(M_Z^2)\simeq-\frac{\alpha}{3\pi}\frac{4}{15}\frac{M_Z^2}{m_t^2}\simeq -6 \times 10^{-5}$.}
\be
\Delta \alpha_{\rm lep} (M_Z^2) \; \simeq \; 314.98 \: \times \: 10^{-4}.
\ee
In contrast, the corresponding free quark loop contribution gets
substantially modified by low energy strong interaction effects, which
cannot be obtained by perturbative QCD (pQCD).  As already mentioned,
fortunately, one can evaluate this hadronic term $\Delta \al _{\rm
had}^{(5)}$ from hadronic $\epm $- annihilation data by using a
dispersion relation. The relevant once subtracted vacuum polarization
amplitude (\ref{VPamp}) satisfies a convergent dispersion relation and
correspondingly the shift of the fine structure constant $\alpha$ is
given by

\ba
\Delta^{(5)} \alpha_{\rm had} = - \frac{\alpha s}{3\pi}\;\bigg(\!\!\!\!\!\!\!\!\! &&{\rm
P}\!\!\!\!\!\!  \int_{4m_\pi^2}^{E^2_{\rm cut}} ds'
\frac{R^{\mathrm{data}}_\gamma(s')}{s'(s'-s)}
+ {\rm P}\!\!\!\!\!\!
\int_{E^2_{\rm cut}}^\infty ds'
\frac{R^{\mathrm{QCD}}_\gamma(s')}{s'(s'-s)}\,\, \bigg)
\ea
where
\be
R_\gamma(s) \equiv \frac{\sigma(e^+e^- \rightarrow \gamma^*
\rightarrow {\rm hadrons})}{ \sigma(e^+e^- \rightarrow \gamma^* \rightarrow
\mu^+ \mu^-)} = 12\pi{\rm Im}\Pi'_{\rm had}(s)\;\;.
\label{Rdef}
\ee

\vspace*{3mm}

\noi Accordingly, the one particle irreducible (1pi) blob

\centerline{%
\includegraphics[scale=0.65]{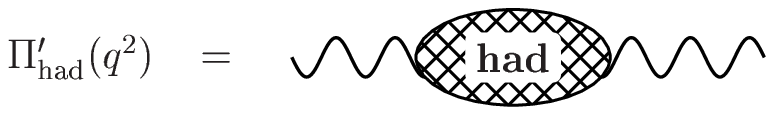}
}

\noi which is the relevant building block in our context and is
given by diagrams which cannot be cut into two disconnected parts by
cutting a single photon line, at low energies exhibits intermediate
states like $\pi^0\gamma,\rho,\omega,\phi,\cdots,\pi\pi,3\pi,4\pi,\cdots,\pi\pi\gamma,\pi\pi Z,$
$\cdots,\pi\pi H,\cdots,KK,\cdots$ (at least one hadron plus any strong,
electromagnetic or weak interaction contribution) and the corresponding
contributions are to be calculated via a dispersion relation from the
imaginary parts which are given by the production of the corresponding
intermediate states in $\epm$--annihilation via virtual photons (at
energies sufficiently below the point where $\gamma-Z$ interference
comes into play).

A direct evaluation of the $R(s)$--data up to $\sqrt{s}=E_{cut}=5$ GeV
and for the $\Upsilon$ resonance--region between 9.6 and 13 GeV and
applying perturbative QCD from 5.0 to 9.6 GeV and for the high energy
tail above 13 GeV at $M_Z=$ 91.19 GeV yields\footnote{pQCD for
calculating $R(s)$, as worked out to high accuracy in
Refs.~\cite{GKL}--\cite{ChHK00}, is used here only where it has been
checked to work and converge well: in non--resonant regions at
sufficiently high energies and sufficiently far from resonances and
thresholds. I have further checked that results obtained with my own
routines agree very well with the ones obtained via the recently
published program {\tt rhad-1.00}~\cite{HS02}.}:
\ba
\label{FJ03M}
\Delta \al _{\rm had}^{(5)}(\mz) &=& 0.027773 \pm 0.000354 \\
\alpha^{-1}(\mz)&=&128.922 \pm 0.049 \nn \;.
\ea
\noi The contributions from different energy ranges are shown in
Tab.~\ref{tab:dal}\footnote{Table 1 also specifies largely details
of the error handling. The different energy ranges mark typical
generation of experiments within which systematic errors are
considered to be 100\% correlated, while all errors are treated as
independent for all entries of the table.}.

\begin{table}[h]
\begin{tabular}{cc||r||r}
\hline
 final state &  energy range (GeV) & $\dahz$ (stat) (syst)
& $\dah0$ (stat) (syst) \\
\hline
$\chi PT$  &    (0.28, 0.32) &    0.04 ( 0.00) ( 0.00)&    0.03 ( 0.00) ( 0.00) \\
$\rho$     &    (0.28, 0.81) &   26.16 ( 0.24) ( 0.27)&   24.26 ( 0.23) ( 0.25) \\
$\omega$   &    (0.42, 0.81) &    3.02 ( 0.04) ( 0.08)&    2.75 ( 0.03) ( 0.07) \\
$\phi$     &    (1.00, 1.04) &    4.74 ( 0.07) ( 0.11)&    4.07 ( 0.06) ( 0.09) \\
$J/\psi$   &                 &   11.50 ( 0.56) ( 0.61)&    4.06 ( 0.19) ( 0.19) \\
$\Upsilon$ &                 &    1.27 ( 0.05) ( 0.07)&    0.07 ( 0.00) ( 0.00) \\
    hadrons&    (0.81, 1.40) &   12.92 ( 0.13) ( 0.52)&   11.05 ( 0.11) ( 0.43) \\
    hadrons&    (1.40, 3.10) &   27.13 ( 0.11) ( 0.60)&   15.75 ( 0.06) ( 0.37) \\
    hadrons&    (3.10, 3.60) &    5.31 ( 0.11) ( 0.10)&    1.90 ( 0.04) ( 0.04) \\
    hadrons&    (3.60, 9.46) &   51.49 ( 0.25) ( 3.00)&    8.41 ( 0.04) ( 0.44) \\
    hadrons&    (9.46,13.00) &   18.59 ( 0.25) ( 1.36)&    0.90 ( 0.01) ( 0.07) \\
    perturb& (13.0,$\infty$) &  115.59 ( 0.00) ( 0.12)&    1.09 ( 0.00) ( 0.00) \\
\hline
    data   &    (0.28,13.00) &  162.14 ( 0.74) ( 3.46)&     73.21 ( 0.33) ( 0.80) \\
    total  &                 &  277.73 ( 0.74) ( 3.46)&     74.30 ( 0.33) ( 0.80) \\
\hline
\end{tabular}
\caption{%
Results for $\dahz^{\mathrm{data}}$ and
$\dah0^{\mathrm{data}}$ ($\sqrt{s_0}=2.5$ GeV).}
\label{tab:dal}
\end{table}

Our analysis is as close to the experimental results as possible by
utilizing the trapezoidal rule together with PDG rules for taking
weighted averages between different experiments as described in detail
in \cite{EJ95}

The most important ingredient of our analysis are the $\epm$--data which
we described in detail in \cite{EJ95} (see also~\cite{ADH98}) and the
new data which have become available since then.  The developments
concerning the experimental data as well as some theoretical aspects are
the following:

\noi ~~$\bullet$ The updated results of the precise measurements of the processes
$e^+e^- \to \rho \to \pi^+\pi^-$, $e^+e^- \to \omega \to
\pi^+\pi^-\pi^0$ and $e^+e^- \to \phi \to K_LK_S$ performed by the
CMD-2 collaboration which have just been presented~\cite{CMD2}. The
update appeared necessary due to an overestimate of the integrated
luminosity in previous analyzes\footnote{This affects in particular
the leading hadronic contribution to the anomalous magnetic moment of
the muon. I now obtain $\amu^{\mathrm{had}(1)}=(695.5\pm
8.6)\:\times\:10^{-10}$ ($\epm$--data based).}. The latter was published in
2002~\cite{CMD}. A more progressive error estimate (improving on
radiative corrections, in particular) allowed a reduction of the
systematic error from 1.4\% to 0.6 \% . Also some other CMD-2 and SND
data at energies $E< 1.4$ GeV have become available and have been
included.

\noi ~~$\bullet$ Before in 2001 BES-II published their final $R$--data which, in the
region 2.0 GeV to 5.0 GeV, allowed to reduce the previously huge systematic
errors of about 20\% to 7\% ~\cite{BES}.

\noi ~~$\bullet$ After 1997 precise $\tau$--spectral functions
became available~\cite{ALEPH,OPAL,CLEO} which, to the extent that
flavor $SU(2)_{\rm f}$ in the light hadron sector is a symmetry, allows to
obtain the iso--vector part of the $\epm$--cross
section~\cite{tsai,eidelman}. This possibility has first been
exploited in the present context in~\cite{ADH98}.

\noi ~~$\bullet$ With increasing precision of the low energy
data it more and more turned out that we are confronted with a serious
obstacle to further progress: in the region just above the
$\omega$--resonance, the iso-spin rotated $\tau$--data, corrected for
the known iso-spin violating effects, do not agree with the
$\epm$--data at the 10\% level~\cite{DEHZ}. Before the origin of this
discrepancy is found it will be hard to make further progress in
pinning down theoretical uncertainties.

\noi ~~$\bullet$ In this context
iso-spin breaking effects in the relationship between the $\tau$-- and
the $\epm$--data have been extensively investigated
in~\cite{CEN}. Whatever uncertainties of the estimated iso-spin
violations might remain, it is very unlikely that they can be made
responsible for the observed discrepancies.

\noi ~~$\bullet$ New results for hadronic $\epm$ cross--sections
are expected soon from KLOE, BABAR and BELLE. These experiments,
running at fixed energies, are able to perform measurements via the
radiative return method~\cite{RR,KLOE,Solodov:2002xu}. Results presented
recently by KLOE seem to agree very well with the final CMD-2
$\epm$--data.

\vspace*{3.3cm}

\begin{figure}[ht]
\begin{picture}(120,60)(5,0)
\includegraphics[scale=0.6]{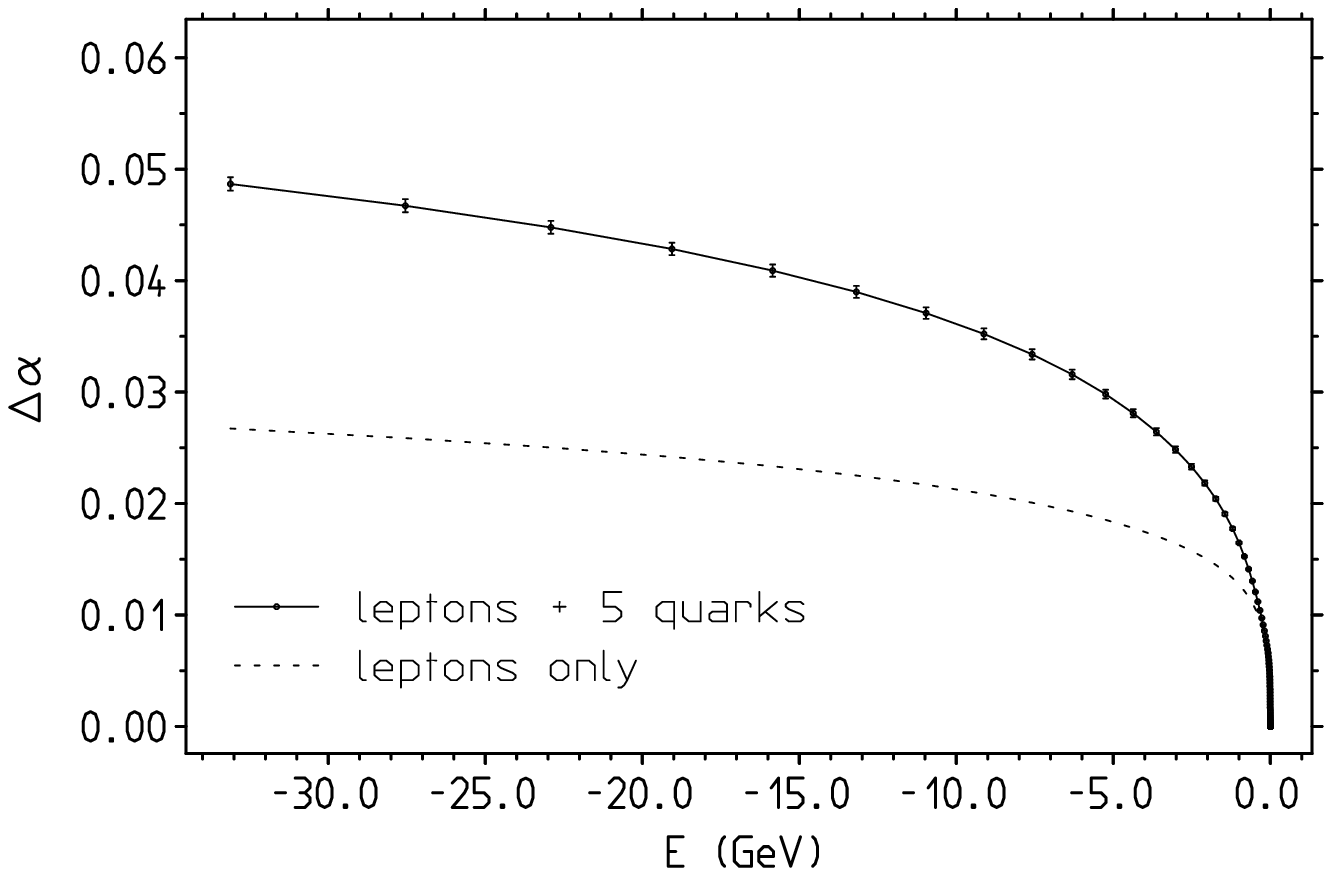}
\end{picture}

\vspace*{-2.0cm}

\begin{picture}(120,60)(-230,0)
\includegraphics[scale=0.6]{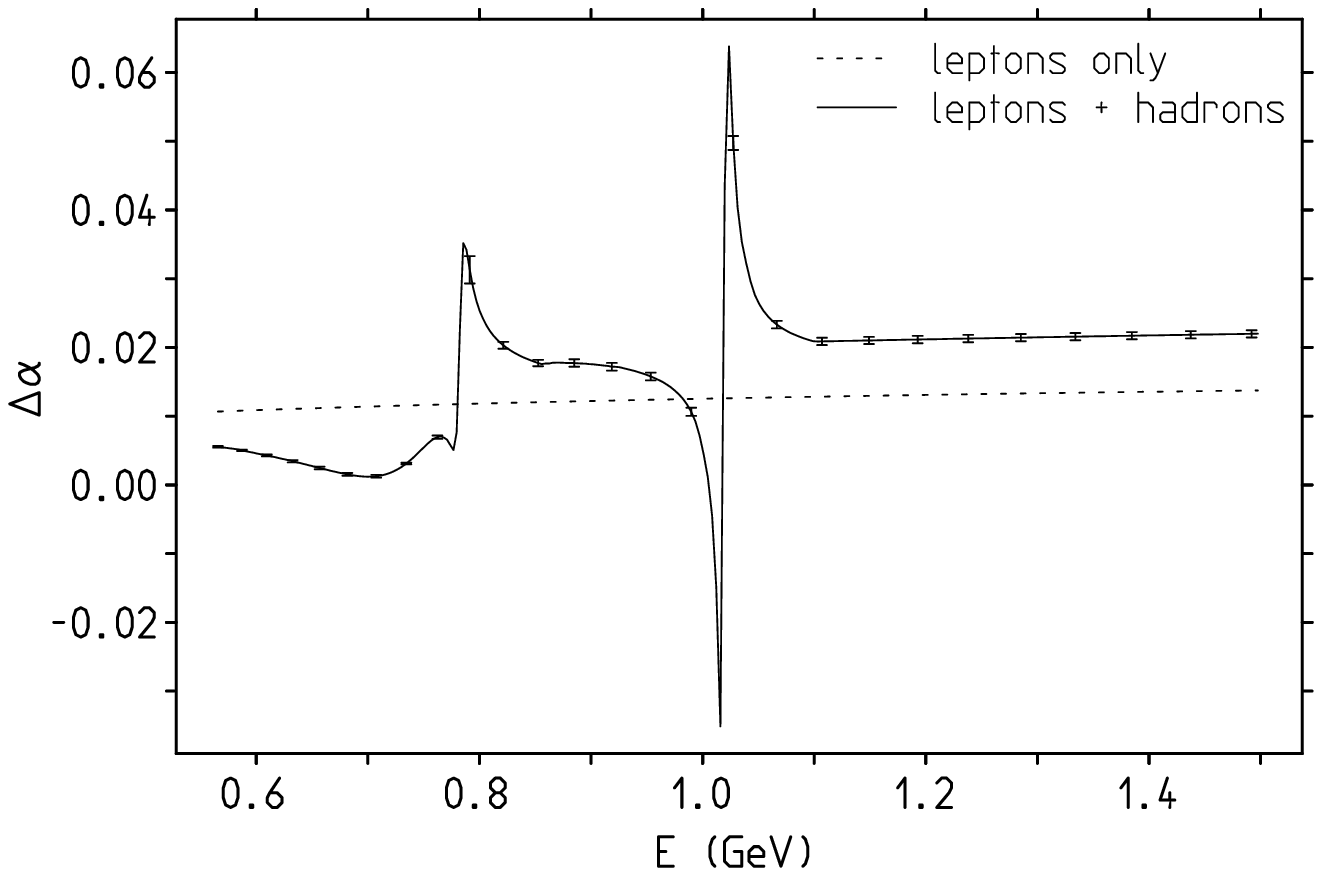}
%{\scalebox{.6 .6}{%
%\epsfbox{alslows.eps}}}
\end{picture}

\vspace*{2mm}

\caption[]{The running of $\alpha$. The ``negative'' $E$ axis is
chosen to indicate space-like momentum transfer. The vertical bars at
selected points indicate the uncertainty. In the time-like region the
resonances lead to pronounced variations of the effective charge
(shown in the $\rho-\omega$ and $\phi$ region).}
\label{fig:alpharun}
\end{figure}
\noi Fig.~\ref{fig:alpharun} illustrates the running of the effective
charge at lower energies in the space-like region.  Typical values
are $\Delta \al (5 \gv) \sim 3\%$ and $\Delta \al (M_Z)
\sim 6\%$, where about $\sim 50\%$ of the contribution comes from
leptons and about $\sim 50\%$ from hadrons.

An analysis similar to ours, however, using piecewise linear approximants
to the non--resonant $R(s)$, which are then integrated analytically,
yields~\cite{BP01}
\ba
\Delta \al _{\rm had}^{(5)}(\mz) &=& 0.027680 \pm 0.000360 \\
\alpha^{-1}(\mz)&=& 128.935 \pm 0.049 \nn \;.
\ea
The precise choice of the lattice used for the linearization
remains somewhat unclear in this method.

An evaluation via the Adler-function which allows to utilize safely
perturbative QCD for the latter at Euclidean energies above 2.5 GeV
(and data at lower energies) yields (see below):
\ba
\label{FJ03E}
\Delta \al _{\rm had}^{(5)}(\mz) &=& 0.027664 \pm 0.000173[0.000137] \\
\alpha^{-1}(\mz)&=&128.937 \pm 0.024 \nn \;.
\ea
The errors here are dominated by the QCD parameter uncertainties and are
given for the worst case correlation [uncorrelated] case.
In contrast to the so called {\em theory--driven} approaches, the Adler
function methods is more elaborate (because it requires to transform
data and theory to the space--like region) but allows for a
dramatically better control of the validity of pQCD.

There are also theory-driven approaches which utilize perturbative QCD
directly for the evaluation of $R(s)$ (and/or for rescaling of the
normalization of the data) and assuming some local version of
quark--hadron duality. A recent result is~\cite{MOR00,HMNT} (see
also~\cite{MZ95}--\cite{deTroconiz:2001yn})
\ba
\Delta \al _{\rm had}^{(5)}(\mz) &=& 0.027690 \pm 0.000180 \\
\alpha^{-1}(\mz)&=&128.933 \pm 0.025 \nn \; .
\ea

\section{$\alpha_{\rm em}(s)$ in precision physics}

A major drawback of the partially non-perturbative relationship
between $\alpha(0)$ and $\alpha(M_Z)$ is that one has to rely on
experimental data exhibiting systematic and statistical errors which
implies a non-negligible uncertainty in our knowledge of the effective
fine structure constant. In precision predictions of gauge boson
properties this has become a limiting factor.  Since $\alpha$, $G_\mu$
and $M_Z$ are the most precisely measured parameters, they are used as
input parameters for accurate predictions of observables like the
effective weak mixing parameter $\sinf$, the vector $v_f$ and
axial-vector $a_f$ neutral current couplings, the $W$ mass $M_W$ the
widths $\Gamma_Z$ and $\Gamma_W$ of the $Z$ and the $W$, respectively,
etc. However, for physics at higher energies, we have to use the
effective couplings at the appropriate scale, for physics at the
$Z$--resonance, for example, $\alpha(M_Z)$ is more adequate to use
than $\alpha(0)$. Of course this just means that part of the higher
order corrections may be absorbed into an effective parameter. If we
compare the precision of the basic parameters
\be \bary{ccccccccccc}
\frac{\delta \alpha}{\alpha} &\sim& 3.6 &\times& 10^{-9}&~~~~~&
\frac{\delta \alpha(M_Z)}{\alpha(M_Z)} &\sim& 1.6 \div 6.8 &\times& 10^{-4}\\
\frac{\delta G_\mu}{G_\mu} &\sim& 8.6 &\times& 10^{-6}&&
\frac{\delta M_Z}{M_Z} &\sim& 2.4 &\times& 10^{-5}
\eary
\ee
we observe that the uncertainty in $\alpha(M_Z)$ is roughly an order
of magnitude worse than the next best, which is the $Z$--mass. Future
TESLA requirements are $\frac{\delta \alpha(M_Z)}{\alpha(M_Z)} \sim 5.3
\times 10^{-5}$ (based on the assumption that a precision $\delta \sin^2 \Theta^\ell_{\rm eff} \simeq
0.000013$ (GigaZ option) and $\delta M_W \sim 6 \mv $ (MegaW) may be
reached)~\cite{TDR}.

Let me remind the reader that $\dal$ enters in electroweak precision
physics typically when calculating versions of the weak mixing
parameter $\sini$ from $\al$, $\Gmu$ and $M_Z$ via
\ba
\sini\:\cosi\: =\frac{\pi \al}{\sqrt{2}\:G_\mu\:M_Z^2}
\frac{1}{1-\Delta r_i}
\ea
where $\Delta r_i =\Delta r_i({\al ,\: \Gmu ,\: M_Z ,}\:m_H,
\:{m_{f\neq t},\:m_t})$ includes the higher order corrections, which can be calculated in the
SM or in alternative models. $\Delta r$ has been calculated for the
first time by A.~Sirlin in 1980~\cite{Sirlin80}. In the SM,
today, the Higgs boson mass $m_H$ is the only relevant unknown
parameter and by confronting the calculated with the experimentally
determined value of $\sini$ one obtains important indirect constraints
on the Higgs mass. $\Delta r_i$ depends on the definition of
$\sini$. The various definitions coincide at tree level and hence only
differ by quantum effects. From the weak gauge boson masses, the
electroweak gauge couplings and the neutral current couplings of the
charged fermions we obtain
\ba
\sinW &=& 1-\frac{M_W^2}{M_Z^2}\\
\sing &=& e^2/g^2=\frac{\pi \al}{\sqrt{2}\:G_\mu\:M_W^2}\\
\sinf &=&
\frac{1}{4|Q_f|}\;\left(1-\frac{v_f}{a_f} \right)\;,\;\;f\neq \nu\;,
\ea
for the most important cases. $\Delta r_i$ usually is written in the form
\ba
\Delta r_i &=& \dal - f_i(\sini)\:\dro + \Delta r_{i\:\mathrm{reminder}}
\ea
with a universal term $\dal$ which affects the predictions of
$M_W$, $A_{LR}$, $A^f_{FB}$, $\Gamma_f$, etc. The uncertainty $\delta
\Delta \alpha$ implies uncertainties $\delta M_W$, $\delta \sini$
given by\footnote{This compares with the second major source of
uncertainty coming from the top mass [currently $\delta m_t/m_t \sim 2.9
\times 10^{-2}$]
\bea
\frac{\delta M_W}{M_W} &\sim& \frac{\cosW}{\cosW-\sinW}
\; \damt \sim 1.426 \;\damt \\
\frac{\delta \sinf}{\sinf} &\sim& ~- 2\:\frac{\cosf}{\cosf-\sinf}
\;\damt \sim -2.852 \; \damt\;.
\eea
}
\ba
\frac{\delta M_W}{M_W} &\sim& \ha \frac{\sinW}{\cosW-\sinW}
\;\delta \dal \sim 0.213 \;\delta \dal \\
\frac{\delta \sinf}{\sinf} &\sim& ~~\frac{\cosf}{\cosf-\sinf}
\;\delta \dal \sim 1.54 \;\delta \dal\;.
\label{deltas}
\ea
The present indirect Higgs mass ``measurement'' reads
$m_H= 96^{+60}_{-38}$ GeV}. The discrepancy between $\epm$-- and
$\tau$--data based evaluations amounts to $\delta m_H \sim - 19 $ GeV~\cite{DEHZ}
(the direct lower bound is $m_H > 114$ GeV at 95\% CL while the
indirect upper bound reads $m_H < 219$ GeV at 95\% CL (1-sided).
For more details we refer to~\cite{MG} (in these proceedings).

\section{The $\tau$ vs. $\epm$ problem}

The iso-vector part of $\sigma(e^+e^- \to {\rm
hadrons})$ may be calculated by an iso-spin rotation from $\tau$--decay
spectra, to the extend that the conserved vector current is conserved
(CVC). The relation may be derived by comparing diagrams like:

\centerline{%
\includegraphics[scale=0.75]{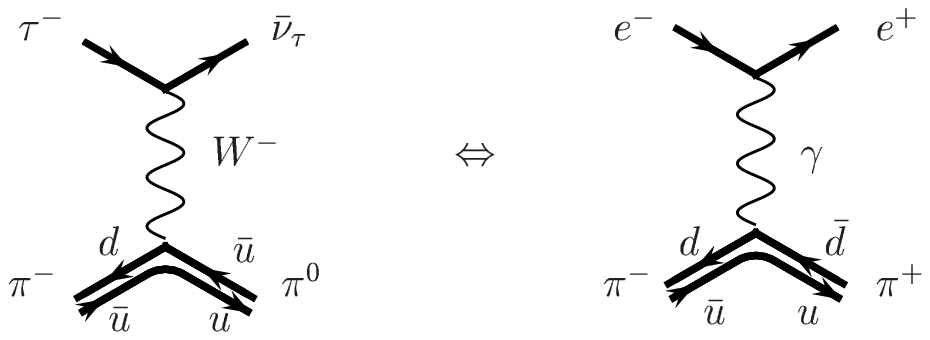}
}

Thus comparing $\tau^- \ra X^- \nu_\tau$ with $e^+ e¯ \ra X^0$ the
hadronic states $X^-$ and $X^0$ are approximately related by an
iso-spin rotation if the states are $I=1$ iso-vector states.  The
$\epm$ cross--section is then given by
\ba
\sigma_{\epm \ra X^0}^{I=1}= \frac{4
\pi \al^2}{s}v_{X^-}\;\;,\;\;\;
\sqrt{s} \leq m_\tau
\ea
in terms of the $\tau$ spectral function $v_V$.
The $\tau$ \sf\ $v_{V}(s)$ for a given vector hadronic state $V$ is
defined by
\ba
   v_V(s)
   \equiv
           \frac{m_\tau^2}{6\,|V_{ud}|^2\,S_{\mathrm{EW}}}\,
              \frac{B(\tau^-\rightarrow \nu_\tau\,V^-)}
                   {B(\tau^-\rightarrow \nu_\tau\,e^-\,\bar{\nu}_e)}
              \frac{1}{N_{V}}\frac{d N_{V}}{ds}\,
              \left[ \left(1-\frac{s}{m_\tau^2}\right)^{\!\!2}\,
                     \left(1+\frac{2s}{m_\tau^2}\right) \right]^{-1},
\ea
where $|V_{ud}|=0.9752\pm0.0007$~\cite{PDG02} denotes the CKM weak
mixing matrix element and $S_{\mathrm{EW}}=1.0233\pm0.0006$ accounts for electroweak
radiative corrections.  The \sfs\ are obtained from
the corresponding invariant mass distributions.

Before a precise comparison is possible all kind of iso-spin breaking
effects have to be taken into account. As mentioned earlier, this has
been investigated in~\cite{CEN} for the most relevant $\pi \pi$
channel. Writing
\ba
\sigma_{\pi\pi}^{(0)}=\left[ \frac{K_{\sigma}(s)}{K_\Gamma(s)}\right]
\: \frac{d \Gamma_{\pi\pi [ \gamma ] }}{ds} \times
\frac{R_{\rm IB}(s)}{S_{\rm EW}}
\ea
with
\ba
K_\Gamma (s) = \frac{G_F^2\:|V_{ud}|^2\:m_\tau^3 }{384 \pi^3}\;\left(1-\frac{s}{m_\tau^2}
\right)^2\; \left( 1+2\: \frac{s}{m_\tau^2}\right)\;\;;\;\;\; K_\sigma
(s) = \frac{\pi \alpha^2}{3 s}\;,
\ea
the iso-spin breaking correction
\ba
R_{\rm IB}(s) = \frac{1}{G_{\rm EM}(s)} \:
\frac{\beta^3_{\pi^+\pi^-}}{\beta^3_{\pi^+ \pi^0}} \: \left| \frac{F_V(s)}{f_+(s)}\right|^2
\ea
include the photonic corrections (based on scalar QED), phase space
corrections due to the $\pi^\pm - \pi^0$ mass difference and the
form--factor corrections which are dominated by the $\rho - \omega$
mixing effects.  These corrections were applied in~\cite{DEHZ} and
were not able to resolve the puzzle of the observed discrepancy (see
~\cite{DEHZ} for details)\footnote{The only large effect I am aware of
(of order 10\%) which is in the game of the comparison is a possible
shift of the invariant mass of the pion-pairs in the $\rho$ resonance
region. An idea one gets if one is looking at the experimental
$\rho$--mass values, shown in the particle data tables~\cite{PDG02}
(``dipole shape'').  If the energy calibration of the $\pi\pi$--system
would be to low in $\epm$ measurements or to high in $\tau$
measurements by 1\% one could easily get a 10\% decrease or increase
in the tail, respectively. Since the $\rho^\pm-\rho^0$ mass difference
as well as the difference in the widths $\Gamma^{\pm,0}(\rho \to
\pi\pi,\pi\pi\gamma)$ are neither experimentally nor theoretically
established, corresponding iso-spin violations cannot be corrected for
appropriately. Note that the subtraction of the large and strongly
energy dependent vacuum polarization effects (see
Fig.~\ref{fig:alpharun}) necessary for the $\epm$--data, which seems
to worsen the problem, is properly treated in the analysis.}.

\section{Controlling pQCD via the Adler function}

In view of the increasing precision LEP experiments have achieved
during the last few years, more accurate theoretical prediction became
desirable. As elaborated in the introduction, one of the limiting
factors is the hadronic uncertainty of $\dahz$. Because of the large
uncertainties in the data, many authors advocated to extend the use of
perturbative QCD in place of
data~\cite{MZ95}--\cite{deTroconiz:2001yn}. The assumption that pQCD
may be reliable to calculate (\ref{Rdef})
down to energies as low as 1.8 GeV seems to be supported by

\noi ~~$\bullet$ the apparent applicability of pQCD to $\tau$ physics. 
In fact the running of $\alpha_s(m_\tau) \ra \alpha_s(M_Z)$ from the
$\tau$ mass up to LEP energies agrees well with the LEP value. The
estimated uncertainty may be debated, however.

\noi ~~$\bullet$ the smallness~\cite{DH98a} (see also:~\cite{FJ86}) of
non--perturbative (NP) effects if parameterized as prescribed by the
operator product expansion (OPE) of the electromagnetic current
correlator~\cite{SVZ}
\ba
\label{NP}
\Pi_\gamma^{'\mathrm{NP}}(Q^2) &=& \frac{4\pi\al}{3} \sum\limits_{q=u,d,s}
Q_q^2 N_{cq}\,
\cdot \bigg[\frac{1}{12}
\left(1-\frac{11}{18}a\right)
\frac{<\frac{\alpha_s}{\pi} G G>}{Q^4} \crn
&+&2\,\left(1+\frac{a}{3} +\left(\frac{11}{2}-\frac34
l_{q\mu}
\right)\,a^2\right)
\frac{<m_q \bar{q}q>}{Q^4} \\
&+& \left(\frac{4}{27}a
+\left(\frac{4}{3}\zeta_3-\frac{257}{486}-\frac13 l_{q\mu}
\right)\,
a^2\right)
\sum\limits_{q'=u,d,s} \frac{<m_{q'} \bar{{q'}}{q'}>}{Q^4}\,\bigg] \crn
&+& \cdots
 \nn
\ea
where $a\equiv \alpha_s(\mu^2)/\pi$
and $l_{q\mu}\equiv\ln(Q^2/\mu^2)$. $<\frac{\alpha_s}{\pi} G G>$ and
$<m_q \bar{q}q>$ are the scale-invariantly defined condensates.\\

Progress in pQCD here comes mainly from~\cite{mqcd3}. In addition an
exact two--loop calculation of the renormalization group (RG) in the
background field \MOM scheme (BF-MOM) is available~\cite{JT98}. This
allows us to treat ``threshold effects'' closer to physics than in the
\MSb scheme. The BF-MOM scheme respects the QCD Slavnov-Taylor
identities (non-Abelian gauge symmetry) but in spite of that is gauge
parameter ($\xi$) dependent\footnote{In applications considered below
all numerical calculations have been performed in the ``Landau gauge''
$\xi=0$.}.

\begin{figure}[h]
\begin{picture}(120,60)(-40,00)
\includegraphics[scale=0.9]{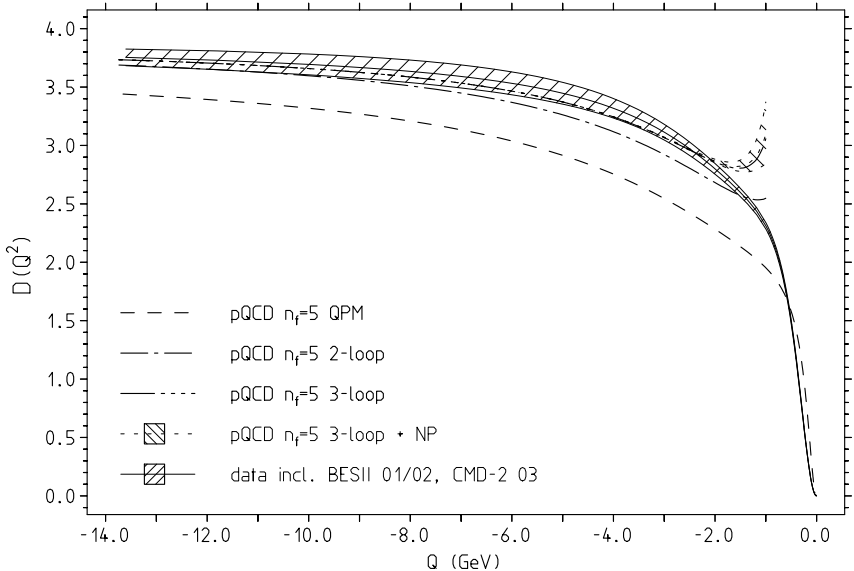}
%\scalebox{.9 .9}{%
%\epsfbox{adleraug03.eps}}
\end{picture}

\caption[]{Adler function: theory vs. experiment~\cite{EJKV98}.}
\label{fig:Adler}
\end{figure}

In Ref.~\cite{EJKV98} a different approach of pQCD improvement was
proposed, which relies on the fact that the vacuum polarization
amplitude $\Pi(q^2)$ is an analytic function in $q^2$ with a cut in
the $s$--channel $q^2=s \geq 0$ at $s\geq 4m^2_\pi$ and a smooth
behavior in the $t$--channel (space-like or Euclidean region). Thus,
instead of trying to calculate the complicated function $R(s)$, which
obviously exhibits non-perturbative features like resonances, one
considers the simpler Adler function in the Euclidean
region. In~\cite{EJKV98} the Adler function was investigated and pQCD
was found to work very well above 2.5 GeV, provided the exact
three--loop mass dependence was used (in conjunction with the
background field \MOM scheme). The Adler function may be defined as a
derivative
\be
D(-s)=-(12\pi^2)\,s\,\frac{d\Pi'_{\gamma}\,(s)}{ds}
=\frac{3\pi}{\alpha} s\frac{d}{ds}\Delta \alpha_{\mathrm{had}}(s)
\label{DD}
\ee
of (\ref{Rdef}) which is the hadronic contribution to the shift of the
fine structure constant. It is represented by
\ba
D(Q^2)=Q^2\:\left(\int_{4 m_{\pi}^2}^{E^2_{\rm cut}}
\frac{  R^{\rm data}(s)}{(s+Q^2)^2}ds\;+\;
\int_{E^2_{\rm cut}}^{\infty}\frac{  R^{\rm pQCD}(s)}{(s+Q^2)^2}ds\:\right)
\label{DI}
\ea
in terms of the experimental $\epm$--data. The standard evaluation
(\cite{EJ95}) of (\ref{DI}) then yields the non--perturbative
``experimental'' Adler function, as displayed in Fig.~\ref{fig:Adler}
for the lower energies where it becomes non--perturbative.

\vspace*{55mm}

%\vspace*{6mm}

For the pQCD evaluation it is mandatory to utilize the calculations with
massive quarks which are available up to three--loops~\cite{mqcd3}.  The
four-loop corrections are known in the approximation of massless
quarks~\cite{GKL}. The outcome of this analysis is pretty surprising and
is shown in Fig.~\ref{fig:Adler}. For a discussion we refer to the
original paper~\cite{EJKV98}. The result was obtained using the
background--field \MOM renormalization scheme, mentioned before. In
the transition from the \MSb to the \MOM scheme we adapt the rescaling
procedure described in~\cite{JT98}, such that for large $\mu$
\bea
\overline{\alpha}_s((x_0\mu)^2)=\alpha_s(\mu^2)+0+ O(\alpha_s^3)\;.
\eea
This means that $x_0$ is chosen such that the couplings coincide
to leading and next--to--leading order at asymptotically large
scales. Numerically we find $x_0\simeq 2.0144$. Due to this
normalization by rescaling the coefficients of the Adler--function
remain the same in both schemes up to three--loops. In the \MOM scheme
we automatically have the correct mass dependence of full QCD, i.e.,
we have automatic decoupling and do not need decoupling by hand and matching
conditions like in the \MSb scheme. For the numerical
evaluation we use the pole quark masses~\cite{PDG02}
$m_c=1.55 \gv ,\;m_b=4.70 \gv,
\;m_t=173.80 \gv \;$ and the strong interaction coupling
$\alpha_{s\; {\MSbm}}^{(5)}(M_Z) = 0.120 \pm 0.003$.
For further details we refer to~\cite{EJKV98}.

\newpage

According to (\ref{DD}), we may compute the hadronic vacuum
polarization contribution to the shift in the fine structure constant
by integrating the Adler function. In the region where pQCD works
fine we integrate the pQCD prediction, in place of the data. We thus
calculate in the Euclidean region~\cite{FJ98}
\be
\Delta\alpha^{(5)}_{\rm had}(-M_Z^2)
=\left[\Delta\alpha^{(5)}_{\rm had}(-M_Z^2) -\Delta\alpha^{(5)}_{\rm
had}(-s_0)\right]^{\mathrm{pQCD}}+ \Delta\alpha^{(5)}_{\rm
had}(-s_0)^{\mathrm{data}}\;\;.
\ee
A save choice is $s_0=(2.5\, \gv)^2$ where we obtain
\ba
\Delta\alpha^{(5)}_{\rm had}(-s_0)^{\mathrm{data}} =0.007430 \pm
0.000087
\ea
from the evaluation of the dispersion integral (\ref{Rdef}).
With the results presented above we find
\ba
\Delta\alpha^{(5)}_{\rm
had}(-M_Z^2) = 0.027626 \pm 0.000087 \pm 0.000149[0.000101]
\ea
for the Euclidean ($t$--channel) effective fine structure constant.
The second error comes from the variation of the pQCD parameters. In
square brackets the error if we assume the uncertainties from
different parameters to be uncorrelated.  The uncertainties coming
from individual parameters are listed in the following table (masses
are the pole masses):

\begin{center}
\begin{tabular}{cccc}
\hline
\hline
parameter & range & pQCD uncertainty & total error \\
\hline
$\alpha_s$ & 0.117 ... 0.123 & 0.000051 & 0.000155 \\
$m_c$      & 1.550 ... 1.750 & 0.000087 & 0.000170 \\
$m_b$      & 4.600 ... 4.800 & 0.000011 & 0.000146 \\
$m_t$      & 170.0 ... 180.0 & 0.000000 & 0.000146 \\
\multicolumn{2}{c}{all correlated} & 0.000149 & 0.000209 \\
\multicolumn{2}{c}{all uncorrelated} & 0.000101 & 0.000178 \\
\hline
\end{tabular}
\end{center}

The largest uncertainty is due to the poor knowledge of the charm
mass.  I have taken errors to be 100\% correlated. The uncorrelated
error is also given in the table.

Comments:

\noi ~~$\bullet$ Contributions to
the Adler function up to three--loops all have the same sign and are
substantial. Four-- and higher--orders could still add up to
non-negligible contribution. An error for missing higher order terms is
not included. The scheme dependence \MSb versus background field
\MOM has been discussed in Ref.~\cite{JT98}.

\noi ~~$\bullet$ The effective fine structure constant in the 
time--like region ($s$--channel), as required for $\epm$--collider
physics, may be obtained from the Euclidean one by adding the
difference
\ba
\Delta=\dahz-\Delta\alpha^{(5)}_{\rm had}(-M_Z^2) =0.000038 \pm
0.000005\,,
\label{delta}
\ea
which may be calculated perturbatively or directly from the
``non--perturbative''\footnote{Since we utilize pQCD for the high
energy tail in the dispersion integral, $\Delta (s)$ for large $s$ is
dominated by the tail and thus in fact is perturbative.} dispersion
integral. It accounts for the $i\pi$--terms $$\ln (-q^2/\mu^2) =
\ln(|q^2/\mu^2|)+i\pi$$ from the logs.

\noi ~~$\bullet$ One may ask the question whether these terms should be
resummed at all, i.e., included in the running coupling.  Usually such
terms tend to cancel against constant rational terms which are not
included in the renormalization group (RG) evolution. It should be
stressed that the Dyson summation (propagator bubble summation) in
general is not a systematic resummation of leading, sub-leading
etc. terms as the RG resummation is.

It is worthwhile to stress here that the running coupling is {\bf not}
a true function of $q^2$ (or even an analytic function of $q^2$) but a
function of the RG scale $\mu^2$. The coupling as it appears in the
Lagrangian in any case must be a constant, albeit a $\mu^2$--dependent
one, if we do not want to end up in conflict with basic principles of
quantum field theory.  The effective identification of $\mu^2$ with a
particular value of $q^2$ must be understood as a subtraction
(reference) point.

\vspace*{55mm}

\begin{figure}[th]
\begin{picture}(120,60)(30,00)
\includegraphics[scale=1.0]{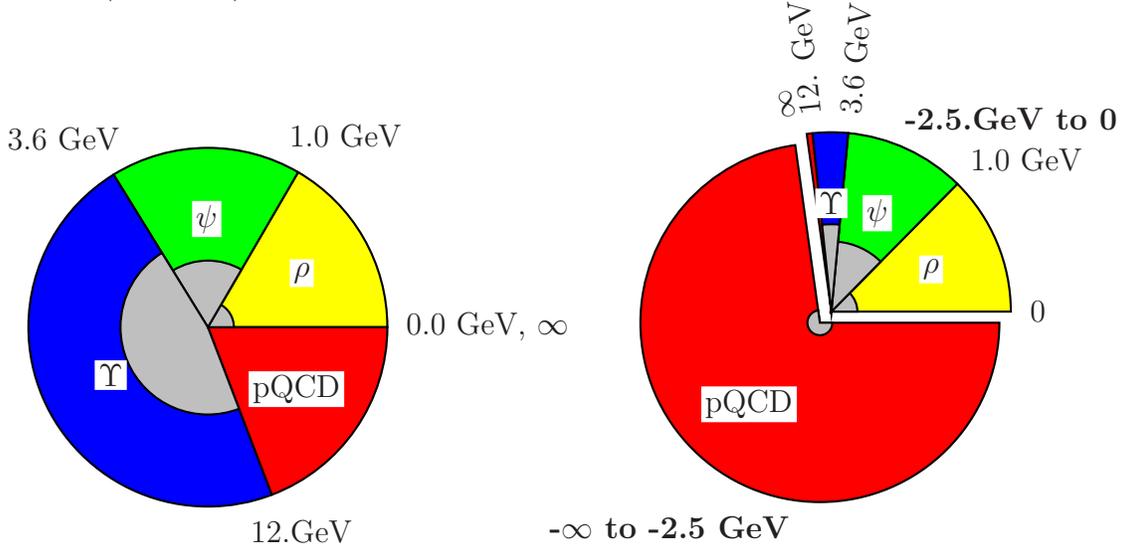}
%\epsfbox{aldistn.eps}
\end{picture}

\caption{Comparison of the distribution of contributions and errors
(shaded areas scaled up by 10) in the standard (left) and the Adler
function based approach (right), respectively.}
\label{fig:alpsta}
\end{figure}

Since $\Delta$ Eq.~(\ref{delta}) is small we may include it in the resummation
without further worrying and thus obtain
\ba
\Delta\alpha^{(5)}_{\rm
had}(M_Z^2) = 0.027664 \pm 0.000173[0.000137] \;\;.
\ea
The alternative evaluation by the Euclidean approach is
compared with the standard evaluation in Tab.~\ref{tab:dal}.
The two methods (standard vs. Euclidean) of evaluating $\dahz$ are
also compared in Fig.~\ref{fig:alpsta}

Our alternative procedure to evaluate $\dahzE$ in the Euclidean region
has several advantages as compared to other approaches used so far:
The virtues of our analysis are the following:

\noi ~~$\bullet$ no problems with the physical threshold and resonances

\noi ~~$\bullet$ pQCD is used only in the Euclidean region and not
below 2.5 GeV.  For lower scales pQCD ceases to describe properly the
functional dependence of the Adler function~\cite{EJKV98} (although the
pQCD answer remains within error bands down to about 1.6 GeV).

\noi ~~$\bullet$ no manipulation of data must be applied and we need 
not refer to global or even local duality. That power corrections of
the type Eq.~(\ref{NP}) are negligible has been known for a long time.
This, however, does not proof the absence of other kind of
non-perturbative effects. Therefore our conservative choice of the
minimum Euclidean energy seems to be necessary.

\noi ~~$\bullet$ According to Tab.~\ref{tab:dal} our non--perturbative
``remainder'' $\dah0$ is mainly sensitive to low energy data, which
changes the chances of possible future experimental improvement
dramatically, as illustrated in Fig.~\ref{fig:alpsta}.

%\newpage

While the uncertainties to $\dahz$ in the standard approach are coming
essentially from everywhere below $M_\Upsilon$, which would make a
new scan over all energies for a precision measurement of
$\sigma_{\mathrm{had}}\equiv \sigma(e^+e^- \rightarrow \gamma^*
\rightarrow {\rm hadrons})$ unavoidable, the new approach leads to a
very different situation. The uncertainty of $\dah0$ is completely
dominated by the uncertainties of data below $M_{J/\psi}$ and thus
new data on $\sigma_{\mathrm{had}}$ are only needed below about 3.6 GeV
which could be covered by a tunable ``$\tau$--charm facility''.

\section{Status and outlook}

Recent result obtained by different authors for the hadronic
contributions to $\alpha_{\rm em}(M_Z)$ are in fairly good
agreement. The estimated uncertainties vary substantially, depending
mainly on additional theoretical assumptions made in the analyzes.
Table~\ref{tab:alperr} compares our results with results obtained by
other authors which obtain smaller errors because they are using pQCD in
a less controlled manner.

\begin{table}[ht]
\begin{tabular}{ll|c|c|cc}
\hline
$\dahz$ &$\delta \dal$ & $\delta \sinf$& $\delta M_W$ & Method & Ref. \\
\hline
0.02800&0.00065~~ & 0.000232 & 11.1 & data $< ~12.~~\gv$         &\cite{EJ95} \\
0.027773 & 0.000354  & 0.000126 &  6.1 &  ~\cite{EJ95} + new data
CMD \& BES      & (\ref{FJ03M}) \\
0.027664&0.000173 & 0.000062 &  3.0 & Euclidean $> ~2.5~~\gv$    & (\ref{FJ03E})\\
0.027680&0.000360  & 0.000128 &  6.2 & data $< ~12.~~\gv$   &\cite{BP01}\\
0.02777&0.00017~~ & 0.000061 &  2.9 & data $< ~1.8~~\gv$         &\cite{KS98}\\
0.02763&0.00016~~ & 0.000057 &  2.7 & data $< ~1.8~~\gv$         &\cite{DH98b}\\
0.027690&0.000180  & 0.000064 &  3.1 & scaled data,
 pQCD 2.8-3.7, 5-$\infty$  &\cite{MOR00,HMNT}\\
~~~~~-&0.00007~~ & 0.000025 &  1.2 &{$\delta \sigma \:\lapprox\: 1\%$ up to $J/\psi$}\\
~~~~~-&0.00005~~ & 0.000018 &  0.9 &{$\delta \sigma \:\lapprox\: 1\%$ up to $\Upsilon$}\\
\multicolumn{2}{c|}{world average} & 0.000170 & 23.0 & LEPEWWG 2003 \\
\hline
\end{tabular}
\caption{%
$\dahz$ and its uncertainties in different evaluations and their
contribution to the errors of $\sinf$ and $ M_W$ according to
(\ref{deltas}). Two entries show what can be reached by increasing the
precision of cross section measurements to 1\%. $\delta M_W$ in MeV.}
\label{tab:alperr}
\end{table}

The reduction of theoretical and experimental uncertainties must go on
to cope with the increased energy and luminosity at future colliders
(like TESLA).  Ideally a reduction of the errors in $\alpha_{\rm
em}(M_Z)$ by about a factor 5 should be achieved. Such progress is
equally important for future precision experiments at lower energies.

In order to be able to rely more on pQCD the required reduction of
errors can be achieved only if the precision of QCD parameters improves
accordingly. On the one hand this proceeds along the traditional
perturbative QCD vs. experimental data line (see~\cite{Kuhn:2001dm} and
references therein), on the other hand, lattice QCD calculations will
become of increasing importance, in this context. A lot has been
achieved in this direction already in recent
years~\cite{DellaMorte:2002vm}-\cite{Lellouch:2002nj}.
%\cite{Rolf:2002gu}%\cite{Knechtli:2002vp}%\cite{Aoki:2002uc}

However, equally important, experimental efforts must go on in measuring
\sigh at the 1\% level up to energies 3.6 GeV. As most of the existing
facilities which are able to measure \sigh have approved or discuss
upgrade programs, we are confident that the progress needed actually
will take place.\\

{\bf Acknowledgments\\~}

It is a pleasure to thank H.~Leutwyler, A.~Stahl, K.~M\"onig and
S.~Eidelman for numerous fruitful discussions.\\

\end{document}